\documentclass[aps,prd,twocolumn,a4paper,superscriptaddress,floatfix]{revtex4-1}

\usepackage{graphicx}
\usepackage{amssymb}
\usepackage{natbib}
\usepackage{color}
\usepackage[dvipsnames]{xcolor}
\usepackage{amsmath}    
\usepackage{mathrsfs}   
\usepackage{hyperref}

\DeclareFontFamily{U}{mathx}{\hyphenchar\font45}
\DeclareFontShape{U}{mathx}{m}{n}{<-> mathx10}{}
\DeclareSymbolFont{mathx}{U}{mathx}{m}{n}

\DeclareMathAccent{\widebar}{0}{mathx}{"73} 

\begin{document}

\newcommand\lsim{\mathrel{\rlap{\lower4pt\hbox{\hskip1pt$\sim$}}
 	\raise1pt\hbox{$<$}}}
\newcommand\gsim{\mathrel{\rlap{\lower4pt\hbox{\hskip1pt$\sim$}}
 	\raise1pt\hbox{$>$}}}

\def\app#1#2{%
  \mathrel{%
    \setbox0=\hbox{$#1\sim$}%
    \setbox2=\hbox{%
      \rlap{\hbox{$#1\propto$}}%
      \lower1.1\ht0\box0%
    }%
    \raise0.25\ht2\box2%
  }%
}
\def\approxprop{\mathpalette\app\relax}

\title{Cosmological Averaging in Nonminimally Coupled Gravity}
\author{S. R. Pinto}
\email{samuel.pinto@astro.up.pt}
\affiliation{Departamento de Física e Astronomia, Faculdade de Ci\^encias, Universidade do Porto, Rua do Campo Alegre, 4169-007 Porto, Portugal}
\affiliation{Instituto de Astrof\'{\i}sica e Ci\^encias do Espa\c co, CAUP, Rua das Estrelas, 4150-762 Porto, Portugal}
\author{P. P. Avelino}
\email{pedro.avelino@astro.up.pt}
\affiliation{Departamento de Física e Astronomia, Faculdade de Ci\^encias, Universidade do Porto, Rua do Campo Alegre, 4169-007 Porto, Portugal}
\affiliation{Instituto de Astrof\'{\i}sica e Ci\^encias do Espa\c co, CAUP, Rua das Estrelas, 4150-762 Porto, Portugal}

\begin{abstract}

We address the challenge, commonly referred to as the cosmological averaging problem, of relating the large-scale evolution of an inhomogeneous universe to that predicted by a homogeneous matter distribution in theories of gravity with nonminimal matter-gravity couplings. To this end, we focus on the class of $f(R,T)$ models given by $f(R,T) = R + F(T)$, where $R$ denotes the Ricci scalar and $T$ the trace of the energy-momentum tensor. This framework provides a simple yet theoretically consistent realization of nonminimal coupled gravity and can be recast as General Relativity minimally coupled to a modified matter Lagrangian. Using global K-monopoles as an illustrative toy model, we show that, when $F$ is a nonlinear function of $T$, the ratio between the spatial average of $F$ and $F$ evaluated at the spatial average of $T$ can deviate significantly from unity and depends on the particle number density. We demonstrate that the common assumption that this ratio is equal to unity generally leads to an inaccurate description of cosmological dynamics. We further show that dust in these theories generally exhibits a non-vanishing proper pressure. Our results highlight the importance of properly accounting for spatial averaging in cosmological models with nonminimal matter-gravity couplings.

\end{abstract}
\date{\today}
\maketitle

\section{Introduction \label{Sec_Intro}}

The Universe is highly inhomogeneous on small and intermediate scales, exhibiting nonlinear structures ranging from particles to planets and stars, and extending to galaxies, clusters, and cosmic filaments \cite{Somerville:2014ika}. By contrast, standard cosmological models \cite{Planck:2018vyg} are based on smooth, homogeneous, and isotropic solutions of Einstein’s equations, described by the Friedmann-Lemaître-Robertson-Walker (FLRW) metric \cite{Robertson:1935zz}. Due to the nonlinear character of Einstein’s equations, it is not guaranteed that the large-scale evolution of an inhomogeneous universe can be well approximated by that of a homogeneous FLRW spacetime. This observation underlies the cosmological averaging, or backreaction, problem, which arises when one attempts to relate the dynamics of an averaged geometry to the average local gravitational dynamics \cite{Buchert:1999er,Buchert:2001sa,Rasanen:2003fy,Kolb:2011zz,Ellis:2011hk,Clarkson:2011zq,Buchert:2019mvq}. In this context, the resulting effective evolution equations generally include additional geometric terms that encode the impact of inhomogeneities. The physical significance and magnitude of geometric backreaction in General Relativity (GR) remain  actively debated (see, e.g., \cite{Baumann:2010tm,Macpherson:2018btl}).

A closely related aspect of the cosmological averaging problem concerns the coarse-graining of the matter sources entering the gravitational field equations. Even if the geometric backreaction effects are small and the spacetime is assumed to be well approximated by an FLRW spacetime on all scales, averaging over fluid inhomogeneities generally yields an effective energy-momentum tensor that does not coincide with that of a homogeneous fluid obeying the same equation of state. Within GR, this issue is largely mitigated due to the minimal coupling between matter and geometry. Nevertheless, even in GR, it can become important in scenarios in which a single component drives both structure formation and cosmic acceleration, such as unified dark energy models \cite{Avelino:2003ig, Beca:2005gc, Avelino:2007tu}, where nonlinear inhomogeneities can substantially modify the effective equation of state.

In this work, we investigate this second aspect of the cosmological averaging problem in modified gravity theories featuring a nonminimal coupling between matter and gravity \cite{Harko:2010mv, Harko:2011kv, Capozziello:2011et, Clifton:2011jh, Alvarenga:2012bt, Alvarenga:2013syu, Haghani:2013oma, Katirci:2013okf, Ludwig_2015, Zaregonbadi:2016xna, Velten:2017hhf, Nojiri:2017ncd, Bahamonde:2017ifa, Avelino:2018rsb, Harko:2018gxr, Azevedo:2018nvi, Minazzoli:2018xjy, Azevedo:2019krx, Fisher:2019ekh, Avelino:2020fek, Rudra:2020nxk, Arruga:2020knb, Azevedo:2021npm, Arruga:2021qhc, Carvalho:2022kxq, Pappas:2022gtt, Goncalves:2023umv, Jana:2023djt, Haghani:2023uad, Solanke:2023aty, Lacombe:2023pmx, Harko:2024sea, Asghari:2024obf, Asghari:2024qgp, Kaczmarek:2024quk, Errehymy:2025kzj, Minazzoli:2025gyw, Boehmer:2025afy, Olmo:2025wot,Chehab:2026obk}. These theories have received increasing attention, motivated by both phenomenological and theoretical considerations, and have been explored as alternatives to a cosmological constant to explain cosmic acceleration. A distinctive feature of these theories is that matter and geometry interact beyond the minimal coupling prescription of GR, leading to modified conservation laws and macroscopic dynamics of fluids and gravity that depend on the physics happening at microscopic scales \cite{Avelino:2018rsb,Ferreira:2020fma}. While these properties open new possibilities, they also introduce conceptual challenges, particularly in the coarse-grained description of matter sources in cosmological applications.

To address the cosmological averaging problem in this class of theories, we focus on $R+F(T)$ gravity as a controlled but illustrative framework. This theory can be reformulated as GR minimally coupled to a modified matter Lagrangian, so that the gravitational sector retains its standard form while all nonminimal matter-geometry interactions are encoded in the matter sector. This makes $R+F(T)$ gravity particularly well suited for isolating the effects of matter coarse-graining and tracing them directly to the underlying nonlinear matter coupling. Moreover, compared to more general nonminimally coupled models, this class of theories is known to evade several of the instabilities identified in the literature \cite{Lacombe:2023pmx} (see also \cite{Harko:2024sea}), while preserving the essential physical features necessary to address the averaging problem. As such, it provides a minimal testbed for assessing the impact of microscopic matter structure on large-scale cosmological dynamics in nonminimally coupled gravity. Furthermore, the cosmological averaging problem has not been taken into account in previous analyses, even within the simpler class of $R+F(T)$ gravity models. Here, we demonstrate that neglecting it can significantly compromise the resulting conclusions.

The outline of this paper is as follows. In Sec. \ref{Sec_f(R,T)}, we briefly review the equations of motion for the gravitational and matter fields in the context of $f(R,T)$ gravity, highlighting the special case of $R + F(T)$ gravity considered in this work. In Sec. \ref{GM}, we study a scalar field multiplet featuring a nonstandard kinetic term and an O(3)-symmetric potential that admits global monopole solutions, known as K-monopoles \cite{Babichev:2006cy,Avelino:2010bu,Pinto:2025plg}. These solutions serve as a particle toy model for exploring various properties of real particles in $R + F(T)$ gravity. Section \ref{Sec_fluids} builds on the results of Sec. \ref{GM} to explore the challenges associated with cosmological averaging for a fluid composed of dust, clarifying several common misconceptions in the literature. Finally, in Sec. \ref{concl}, we summarize our findings and highlight the associated challenges.

Throughout this work, we adopt the metric signature $[-, +, +, +]$ and use natural units where $c = 16\pi G =1$. The Einstein summation convention is employed, meaning that when an index variable appears twice in a single term it implies summation over all possible values of the index. Unless explicitly stated otherwise, Greek indices run over $\{0, 1, 2, 3\}$, while Latin indices are restricted to $\{1, 2, 3\}$.

\section{$f(R,T)$ gravity\label{Sec_f(R,T)}}

Here, we begin by reviewing the general framework of $f(R,T)$ gravity \cite{Harko:2011kv}, which extends GR by allowing the gravitational action to depend explicitly on both the Ricci scalar $R \equiv g^{\mu\nu}R_{\mu\nu}$ and the trace $T \equiv g^{\mu\nu} T_{\mu\nu}$ of the energy–momentum tensor, providing a natural setting to explore the effects of nonminimal matter–gravity couplings.

The action for $f(R,T)$ gravity is given by
\begin{equation}
S = \int d^4x \sqrt{-g} \left[ f(R, T) + \mathcal{L}_{\rm m} \right]\,,
\end{equation}
where $g$ is the determinant of the metric $g_{\mu\nu}$ and $\mathcal{L}_{\rm m}$ is the Lagrangian of the matter fields. The corresponding equations of motion for the gravitational field are given by
\begin{equation}
2(R_{\mu\nu} - \Delta_{\mu\nu}) f_{,R} - g_{\mu\nu} f = \mathcal{T}_{\mu\nu}\,,
\label{Eqm_g_1}
\end{equation}
where a comma denotes a partial derivative. $R_{\mu\nu}$ is the Ricci tensor, $\Delta_{\mu\nu} \equiv \nabla_{\mu} \nabla_{\nu} - g_{\mu\nu} \Box$, $\Box \equiv \nabla_{\mu} \nabla^{\mu}$ and, 
\begin{eqnarray}
\mathcal{T}_{\mu\nu} &\equiv& T_{\mu\nu}  - 2 f_{,T} (T_{\mu\nu} + \mathbb{T}_{\mu\nu})\,, \\
T_{\mu\nu} &\equiv& -\frac{2}{\sqrt{-g}} \frac{\delta (\sqrt{-g} \mathcal{L}_{\rm m})}{\delta g^{\mu\nu}} = g_{\mu\nu} \mathcal{L}_{\rm m} - 2 \frac{\delta \mathcal{L}_{\rm m}}{\delta g^{\mu\nu}}\,, \label{eq_T}\\
\mathbb{T}_{\mu\nu} &\equiv& g^{\alpha\beta} \frac{\delta T_{\alpha\beta}}{\delta g^{\mu\nu}} = \frac{\delta T}{\delta g^{\mu\nu}} - T_{\mu\nu} \,. \label{def_mathbbT}
\end{eqnarray}

Equation \eqref{Eqm_g_1} can be rearranged as
\begin{equation}
G_{\mu\nu}\equiv R_{\mu\nu} - \frac{1}{2} g_{\mu\nu} R =\frac{1}{2} \mathscr{T}_{\mu\nu}\,,
\label{Eqm_g_2}
\end{equation}
where $G_{\mu\nu}$ are the components of the Einstein tensor and 
\begin{equation}
\mathscr{T}_{\mu\nu}=\frac{1}{f_{,R}}\left[2\Delta_{\mu\nu}f_{,R}-g_{\mu\nu}(Rf_{,R}-f)+\mathcal{T}_{\mu\nu}\right]\,.
\label{mathcalT}
\end{equation}
In $f(R,T)$ gravity the standard energy-momentum tensor $T_{\mu\nu}$ is typically not conserved. However, since the Einstein tensor is covariantly conserved, the tensor $\mathscr{T}_{\mu\nu}$ must also satisfy 
\begin{equation}
\nabla^\mu \mathscr{T}_{\mu\nu} = 0\,.\label{Tmodcons}
\end{equation}
However, the physical interpretation of Eq. \eqref{Tmodcons}, is not straightforward since $\mathscr{T}_{\mu\nu}$ is nontrivially dependent on the Ricci scalar and its derivatives. 

\subsection{$R+F(T)$ gravity}

For certain functional forms of $f(R,T)$, the physical meaning of the field equations becomes more transparent. A particularly illustrative example is the separable case, 
\begin{equation}
f(R,T) = f_1(R) + f_2(T)\,. \label{f1pf2}
\end{equation}
In this scenario, the theory can be reformulated as an $f_1(R)$ gravity model with an effective modification of the matter sector. More precisely, the $f_2(T)$ term can be absorbed into the matter Lagrangian, yielding a modified matter Lagrangian
\begin{equation}
\mathscr{L}_{\rm m}=\ \mathcal{L}_{\rm m} + f_2(T)\,,
\end{equation}
which in turn leads to a well-defined modified energy-momentum tensor. In this form, deviations from standard $f(R)$ gravity are completely encoded in the modified matter dynamics induced by the explicit dependence on the trace $T$ of the energy-momentum tensor. 

Although extending the analysis presented in this work to the broader family of models given by Eq.~\eqref{f1pf2} is straightforward, for simplicity we focus on the subclass $f(R,T)=R+F(T)$. These can be recast as GR with a modified matter Lagrangian,
\begin{equation}
\mathscr{L}_{\rm m} = \mathcal{L}_{\rm m} + F \, .
\label{mathscrL}
\end{equation}
For this class of models, $\mathscr{T}_{\mu\nu}$ are the components of a modified energy-momentum tensor given by 
\begin{eqnarray}
\mathscr{T}_{\mu\nu} &\equiv& -\frac{2}{\sqrt{-g}} \frac{\delta (\sqrt{-g} \mathscr{L}_{\rm m})}{\delta g^{\mu\nu}} \nonumber \\ 
&=& T_{\mu\nu} + F g_{\mu\nu} - 2 F_{,T} \left(T_{\mu\nu} + \mathbb{T}_{\mu\nu}\right)\,.
\label{mathcalT2}
\end{eqnarray}
Its trace is then 
\begin{equation}
\mathscr{T} = T + 4 F - 2 F_{,T} \left(T + \mathbb{T}\right)\,.
\label{mathcalT3}
\end{equation}

$R + F(T)$ gravity is dynamically equivalent to GR with a modified matter Lagrangian, which might suggest that it offers little new physical insight \cite{Fisher:2019ekh}. Nevertheless, this equivalence is particularly useful: it provides a familiar and well-controlled framework for exploring the phenomenology of $f(R,T)$ gravity and, more generally, of theories featuring nonminimal couplings between geometry and matter fields. By working in a setting where the geometric sector remains standard while the matter sector is modified in a precise and tractable manner, one can isolate and analyze some of the physical consequences of such couplings without the additional complications associated to nontrivial modifications of the gravitational action itself.

\subsection{Scalar-field matter source \label{Sec_sfm}}

To set the stage for the study of global K-monopoles in the next section, here we briefly review a number of relevant results for a scalar-field matter source.

Consider a matter field described by a real scalar field multiplet $\phi^a$ with Lagrangian
$\mathcal{L}_{\rm m}(\phi^a, X)$, where
\begin{equation}
X = -\frac{1}{2}\,\nabla_\mu \phi^a \nabla^\mu \phi^a \, .
\end{equation}
The corresponding energy-momentum tensor is
\begin{equation}
T_{\mu\nu}
= \mathcal{L}_{{\rm m},X}\, \nabla_\mu \phi^a \nabla_\nu \phi^a
  + \mathcal{L}_{\rm m}\, g_{\mu\nu} \, ,
\label{EM1}
\end{equation}
with trace
\begin{equation}
T \equiv {T^{\mu}}_{\mu}
= -2 X\, \mathcal{L}_{{\rm m},X}
  + 4\, \mathcal{L}_{\rm m} \, .
  \label{Tracephi}
\end{equation}

Noting that
\begin{eqnarray}
\frac{\delta X}{\delta g^{\mu\nu}}
&=& -\frac{1}{2}\,\nabla_\mu \phi^a \nabla_\nu \phi^a \, ,\\
T_{,X}
&=& -2\,(X\,\mathcal{L}_{{\rm m},XX}- \mathcal{L}_{{\rm m},X}) \,,
\end{eqnarray}
one immediately obtains
\begin{equation}
\frac{\delta T}{\delta g^{\mu\nu}}
= \left( X\,\mathcal{L}_{{\rm m},XX}
       - \mathcal{L}_{{\rm m},X} \right)
  \nabla_\mu \phi^a \nabla_\nu \phi^a \, .
\end{equation}

Using Eq.~\eqref{def_mathbbT}, one then finds
\begin{equation}
\mathbb{T}_{\mu\nu}
= \left( X\,\mathcal{L}_{{\rm m},XX}
       - 2\mathcal{L}_{{\rm m},X} \right)
  \nabla_\mu \phi^a \nabla_\nu \phi^a
  - \mathcal{L}_{\rm m}\, g_{\mu\nu} \, , \label{TT}
\end{equation}
whose trace is equal to
\begin{equation}
\mathbb{T}
= -2 X^2 \,\mathcal{L}_{{\rm m},XX}  + 4 X \mathcal{L}_{{\rm m},X} - 4\mathcal{L}_{\rm m} \,. \label{TTracephi}
\end{equation}

The equation of motion for the scalar field multiplet in $R+F(T)$ gravity is given by:
\begin{eqnarray}
    \nabla_\mu\phi^a\left[\mathscr{L}_{{\rm m},X\phi^b}\nabla^\mu \phi^b+\mathscr{L}_{{\rm m},XX}\nabla^\mu X\right] &+&\\
    \mathscr{L}_{{\rm m},\phi^a}+\mathscr{L}_{{\rm m},X}\Box\phi^a&=&0\,, \nonumber
\label{Equation_motion_multiplet}
\end{eqnarray}
with $\mathscr{L}_{\rm m}=\mathcal{L}_{\rm m} +F(T)$.

\section{Global K-monopoles in $R+F(T)$ gravity \label{GM}}

In this section, we consider global K-monopoles as a particle toy model, using them to illustrate various properties of real particles in $R+F(T)$ gravity, assuming that the spacetime metric is locally flat and can be approximated as Minkowskian inside the particle, an approximation generally expected to be extremely accurate at the particle level.

Consider a real scalar field multiplet $\{ \phi^1, \phi^2, \phi^3 \}$, in a $3+1$ dimensional spacetime, described by the Lagrangian
\begin{equation}
\mathcal{L}_{\rm m} = K(X) - V(\phi^a)\,,
\label{eq1}
\end{equation}
with
\begin{equation}
V(\phi^a) = \frac{\lambda}{4} (\phi^a\phi^a-\eta^2)^2\,,
\end{equation}
Here, $\lambda > 0$ is a dimensionless real constant, and $\phi^a\phi^a =\eta^2$ defines the vacuum manifold as a 2-sphere. For simplicity, we use units where $\eta = 1$ throughout the paper. Combined with the choice $c = 16\pi G = 1$ introduced earlier, this completely fixes the system of units used throughout this paper. Although we shall retain $\lambda$ explicitly in all equations, all numerical results presented in this paper will be obtained considering $\lambda = 1$.

Derrick’s theorem states that scalar field theories with standard kinetic and potential terms do not admit nontrivial static solutions with finite energy in more than one spatial dimension \cite{Hobart:1963elz,Derrick:1964ww}. Here, we circumvent this limitation by introducing a nonstandard kinetic term $K(X)$, which enables the existence of static, finite-energy solutions. For the sake of definiteness, we shall assume that 
\begin{equation}
K(X)=X|X|^{\alpha-1}\,,
\end{equation}
with $\alpha \geq 1$. This particular choice for the kinetic term then implies that
\begin{eqnarray}
X\mathcal{L}_{{\rm m}, X}&=&\alpha K(X)\,, \\ 
X \mathcal{L}_{{\rm m}, XX}&=&(\alpha-1)\mathcal{L}_{{\rm m}, X}\,. \label{LMXX}
\end{eqnarray}

To find maximally symmetric static solutions of global K-monopoles, we use the ansatz:
\begin{equation}
\phi^a = \frac{x^a}{r} f(r)\,, 
\label{phi}
\end{equation}
where $r^2 = x^a x^a$, and $f(r)$ is a monotonically increasing radial profile satisfying $f(0)=0$ and $f(\infty)=1$. Note that Eq.~\eqref{phi} implies $ \phi^a \phi^a = f^2(r)$, so that the potential can be expressed as $V(\phi^a)=V(f) = \lambda (f^2-1)^2/4$. Consequently, the minima of $V$ occur at $f=1$. 

The field equation given in Eq.~\eqref{Equation_motion_multiplet}, for a static spherically symmetric K-monopole then becomes 
\begin{widetext}
\begin{equation}
\left[\frac{1+4F_{,T}}{\alpha |X|^{\alpha -1}}+4(4-2\alpha)F_{,TT}f'^2\right]\frac{dV}{df}-\alpha (4-2\alpha)^2F_{,TT}  |X|^{\alpha-1}X'f'= (1+F_{,T}(4-2\alpha))\left[\frac{\alpha-1}{X} X'f' + f''+\frac{2}{r}f'-\frac{2}{r^2}f \right]\,,
\label{Eq_motion_f}
\end{equation}   
\end{widetext}
where now
\begin{eqnarray}
X \equiv X(r) &=& -\frac12\left(f'^2 +\frac{2}{r^2} f^2\right)\,, \label{eq8}\\
X'\equiv X'(r) &=&  -f'f''-2\frac{ff'r^2-rf^2}{r^4}\, \label{eq9}\\
\frac{dV}{df}&=&\lambda f(f^2-1)\,, \label{eq11}
\end{eqnarray}
and a prime denotes differentiation with respect to $r$.

Equations. \eqref{mathscrL}, \eqref{mathcalT2}, \eqref{EM1}, and \eqref{TT} imply that the proper energy density $\rho$ and proper modified energy density $\varrho$ of a static global monopole in a Minkowski spacetime are given by
\begin{eqnarray}
\rho(r)&\equiv&-{T^{0}}_{0}={\mathbb{T}^{0}}_{0} = -\mathcal{L}_{\rm m}  \label{eq12}\,,\\
\varrho(r) &\equiv& -{\mathscr{T}^{0}}_{0}= -\mathscr{L}_{\rm m}=\rho - F \,.
\label{varrho_GM}
\end{eqnarray}
On the other hand, using also Eqs. \eqref{mathcalT3}, \eqref{Tracephi}, \eqref{TTracephi}, and \eqref{LMXX} one can show that the proper pressure, $p$, and proper modified pressure $\mathscr{P}$ are given by
\begin{eqnarray}
p(r) &\equiv& \frac{T{^i}_{i}}{3}= \frac{T-T{^0}_{0}}{3}=-\frac23 X\mathcal{L}_{{\rm m},X} + 
\mathcal{L}_{\rm m}
\label{pp}\,,\\
\mathscr{P}(r) &=& \frac{{\mathscr{T}^i}_{i}}{3}=\frac{\mathscr{T}-{\mathscr{T}^0}_{0}}{3} \nonumber \\
&=&  p + F- 2 F_{,T} (\alpha -2)(\rho +p)\,.
\label{mathcalP}
\end{eqnarray}

The proper modified mass (or energy) of the monopole $\mathscr{M}$ is then
\begin{equation}
\mathscr{M} =4\pi \int_0^\infty \varrho r^2 dr =m-4\pi\int_0^\infty F(T) r^2 dr\,,
\end{equation}
where $m$ is the proper mass defined by
\begin{equation}
m=4\pi \int_0^\infty \rho r^2 dr \,.
\end{equation}
As shown in \cite{Pinto:2025plg}, $m$ is finite only for $\alpha > 3/2$.

\subsection{Numerical Methods}

Before constructing numerical solutions for $f(r)$, we must establish the physical constraints that define a valid global K-monopole. We prioritize solutions that retain the qualitative behavior found in GR, specifically requiring that the proper standard and modified energy densities remain positive throughout the spatial domain and that both the proper standard and modified masses remain finite. Consequently, in addition to the asymptotic limits $f(0)=0$ and $f(\infty)=1$, we impose that $f(r)$ is strictly monotonic. These conditions make the existence of global K-monopole solutions sensitive to the functional form of $F(T)$ and the parameter $\alpha$. To ensure the existence of well-behaved solutions, we first analyze the field equations in the asymptotic regime where $r\gg1$. 

Consider the family of functions $F(T)=\epsilon T$. For $r\gg1$, the field equation [Eq. \eqref{Eq_motion_f}] can be approximated as
\begin{equation}
\frac{1+4F_{,T}}{\alpha |X|^{\alpha -1}}\lambda f(f^2-1)+(1+F_{,T}(4-2\alpha))\frac{2}{r^2}f=0\,.
\label{fieldEq_approx_1}
\end{equation}
Using that $|X| \sim f^2/r^2$ and $F_{,T}=\epsilon$, the solution to Eq. \eqref{fieldEq_approx_1} at leading order is
\begin{equation}
    f(r\gg1) =1-\alpha\frac{1+\epsilon(4-2\alpha)}{\lambda(1+4\epsilon)}r^{-2\alpha}\,.
    \label{f_limit_1}
\end{equation}
To satisfy the monotonicity condition, the coefficient of the $r^{-2\alpha}$ term must be negative. This requirement imposes constraints on the allowed values of $\epsilon$ and $\alpha$, which split into three distinct parameter regimes:
\begin{equation}
\begin{cases}
 \epsilon>0  &{\land} \ \ \  2\alpha <4+\epsilon^{-1}\,, \\
-\frac{1}{4}<\epsilon<0  &{\land}  \ \ \ 2\alpha >4+\epsilon^{-1}\,, \\
\epsilon<-\frac{1}{4}  &\land  \ \ \ 2\alpha <4+\epsilon^{-1}\,.
\end{cases}
\end{equation}

We now consider the family $F(T)=\epsilon |T|^\beta$, with $\beta>1$. In this case, $F_{,T}$ scales as $r^{-2\alpha(\beta-1)}$ and becomes subdominant in Eq. \eqref{fieldEq_approx_1}. Consequently, in the tail region of the global K-monopole ($r \gg 1$), the field equation reduces to
\begin{equation}
\frac{1}{\alpha|X|^{\alpha-1}}\lambda f(f^2-1)+\frac{2}{r^2}f = 0\,,
\label{fieldEq_approx_2}
\end{equation}
yielding the solution:
\begin{equation}
f(r) = 1 - \frac{\alpha}{\lambda} r^{-2\alpha}\,.
\label{f_limit_2}
\end{equation}

Given the established parameter constraints and asymptotic solutions, we solve the full second-order differential equation [Eq.~\eqref{Eq_motion_f}] by combining numerical integration with the corresponding analytical expressions for $f(r)$. In the near-core region, we employ a shooting method (see, e.g., \cite{Press2007}), imposing boundary conditions near the monopole center ($r=r_0\sim0$), where the field behaves as $f(r\ll1)\sim Ar$, and integrating outward to a much larger radius $r=r_1$, where $f(r)$ is already expected to be extremely close to unity. The shooting parameter $A$ is iteratively adjusted using a binary search algorithm designed to ensure that $f(r_1)$ approaches unity. The final solution $f(r)$ is obtained by smoothly matching the numerical core solution to the asymptotic tail. For definiteness, we set $\lambda=1$ in the numerical analysis, a choice that does not affect our main results.

\subsection{Von Laue condition \label{k-monopoles consequences}}

 In this subsection we demonstrate that the spatial average of the proper pressure does not generally vanish in $R + F(T)$ gravity, signaling a violation of the standard von Laue condition \cite{Laue:1911,Avelino:2018qgt,Giulini:2018tuw,Avelino:2023rac,Pinto:2025plg}. Although this was shown in \cite{Avelino:2024pcl} for particle-like solutions in  $1+1$ dimensions, here we generalize the result to $3+1$ dimensions using global K-monopoles as a particle toy model.

\begin{figure}[t!]
\centering
\includegraphics[width=\columnwidth]{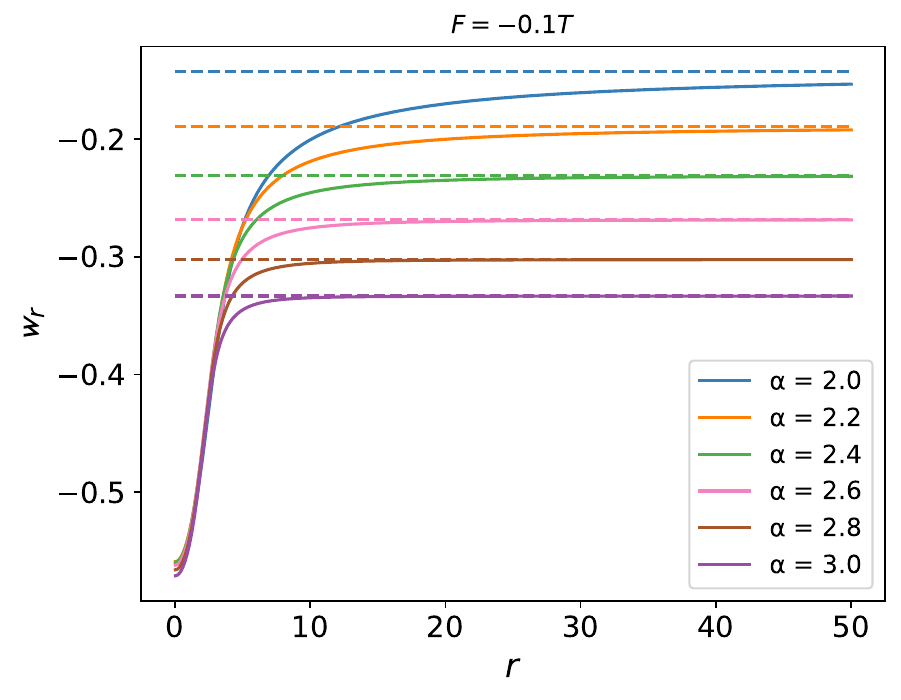}
\caption{The solid lines show the value of $w_ r\equiv \langle p \rangle_r / \langle \rho \rangle_r$ as a function of the distance to the monopole center $r$ for static global K-monopole solutions obtained numerically considering various values of $\alpha$ and $F(T) = -0.1T$. The colored horizontal dashed lines represent the corresponding asymptotic limits $w_\infty$, defined in Eq. \eqref{f(R,T):barw_infty} for each value of $\alpha$.
\label{fig1}
}
\end{figure}

Static global K-monopoles of finite mass satisfy a modified von Laue condition \cite{Avelino:2024pcl}, which in cartesian coordinates can be expressed as:
\begin{equation}
\int_\Sigma \mathscr{T}^{\mu i}\, d^3x = 0 \, ,
\label{von laue}
\end{equation}
where $\Sigma$ denotes the spatial volume of Minkowski space at a fixed physical time $t$. Using Eq. \eqref{von laue} together with the expression for the proper modified pressure $\mathscr{P}$ [Eq. \eqref{mathcalP}], one finds that
\begin{eqnarray}
\int \mathscr{P}\, d^3x &=&  \int \left[p + F- 2 F_{,T} (\alpha -2)(\rho+p)\right] d^3x \nonumber \\
&=&0\,. \label{f(R,T):int_mathcalP}
\end{eqnarray}
Therefore, the spatial average of the proper pressure $p$ generally does not vanish unless $F = 0$. 

\begin{figure}[t!]
\centering
\includegraphics[width=\columnwidth]{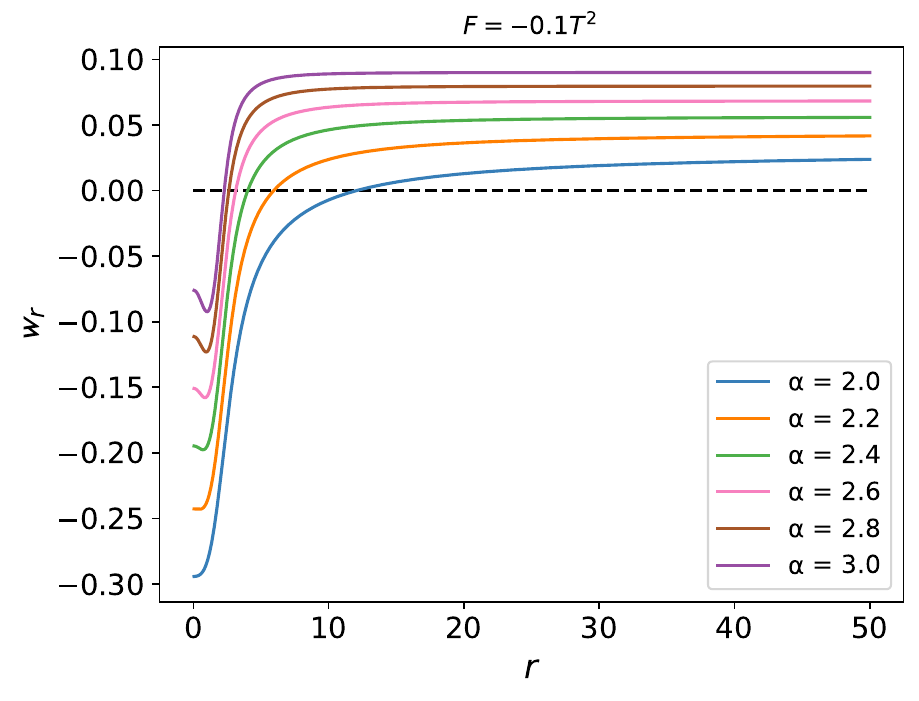}
\caption{The solid lines display the behavior of $w_r$ as a function of the radial coordinate $r$, considering various values of $\alpha$ and $F(T) = -0.1T^2$. The dashed black line indicates the asymptotic value implied by the standard von Laue condition ($w_\infty = 0$), which is not satisfied by any of models considered.
\label{fig2}
}
\end{figure}

For $F(T) = \epsilon T$, the proper modified pressure $\mathscr{P}$ within a global K-monopole can be written in terms of $p$ and $\rho$ as
\begin{equation}
   \mathscr{P} = p\bigl(1 + \epsilon(7 - 2\alpha)\bigr) - \rho\, \epsilon(2\alpha - 3)\,.
   \label{mathcalP(p,rho)}
\end{equation}
Substituting this expression into Eq.~\eqref{f(R,T):int_mathcalP}, we obtain an expression for the volume integral of the proper pressure:
\begin{equation}
\int_0^\infty p\, r^2 \, dr =
\frac{\epsilon(2\alpha - 3)}{1 + \epsilon(7 - 2\alpha)}
\int_0^\infty \rho\, r^2 \, dr\,.
\label{EOS_1}
\end{equation}

For simplicity of notation, we define
\begin{equation}
{w}_r\equiv\langle p\rangle_r/\langle \rho\rangle_r  \,,
\end{equation}
where,
\begin{equation}
\langle \mathcal V \rangle_r = \frac{3 \int_0^r \mathcal V(r') r'^2 dr'}{r^3} \label{average_r}
\end{equation}
represents the spatial average of the physical variable $\mathcal V$ within a sphere of radius $r$ centered on the global K-monopole. Using Eq. \eqref{EOS_1}, the asymptotic expression for $w_r$ is given by 
\begin{equation}
    {w}_\infty \equiv \lim_{r \to \infty} w_r = \frac{\epsilon(2\alpha-3)}{1+\epsilon(7-2\alpha)}\,,
\label{f(R,T):barw_infty}
\end{equation}
which vanishes only in the GR limit ($\epsilon=0$), as required by the standard von Laue condition, or for $\alpha=3/2$.

Figure~\ref{fig1} shows the value of $w_r$ as a function of the distance from the
monopole center, $r$, for static global K-monopole solutions obtained numerically,
assuming $F(T) = -0.1\,T$ and considering several values of $\alpha$ (solid lines). For each value of $\alpha$, $w_r$ approaches a distinct asymptotic value $w_\infty$ at large $r$, as indicated by the corresponding dashed lines and given by Eq.~\eqref{f(R,T):barw_infty}. This behavior demonstrates that the standard von Laue condition does not generally hold in $R + F(T)$ gravity, even when the spacetime metric is assumed to be locally Minkowskian. A qualitatively similar behavior is observed in Fig.~\ref{fig2} for $F(T) = -0.1\,T^2$, with the ratio between the average proper pressure and energy density also asymptotically approaching a constant for $r \gg 1$.

\begin{figure}[t!]
\centering
\includegraphics[width=\columnwidth]{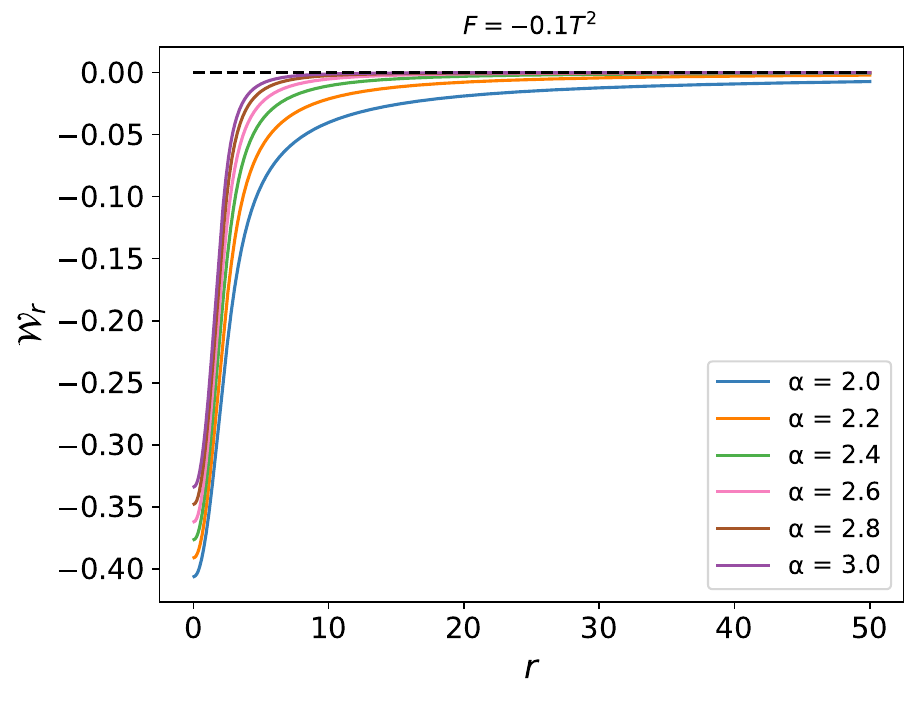}
\caption{The solid lines display the behavior of ${\mathcal W}_r \equiv \langle \mathscr{P} \rangle_r / \langle \varrho \rangle_r$ as a function of the radial coordinate $r$, considering various values of $\alpha$ and $F(T) = -0.1T^2$. indicates the asymptotic value implied by the modified von Laue condition (${\mathcal W}_\infty = 0$). Note that for all models considered (with $\alpha > 3/2$) $\mathcal W_r$ approaches zero at large distances from the global K-monopole center in agreement with the modified von Laue condition.
\label{fig3}
}
\end{figure}

Figure \ref{fig3} shows the dependence of 
\begin{equation}
{{\mathcal W}}_r \equiv \frac{\langle \mathscr{P} \rangle_r}{ \langle \varrho \rangle_r}  \,,
\end{equation}
on the radial coordinate $r$ for several values of $\alpha$ (solid lines), assuming $F(T) = -0.1T^2$. As illustrated, $\mathcal W_r$ approaches zero (dashed black line) for sufficiently large values of $r$ for all models considered with $\alpha > 3/2$. This shows that a modified von Laue condition defined by 
\begin{equation}
{\mathcal W}_\infty \equiv \langle \mathscr{P} \rangle_\infty / \langle \varrho \rangle_\infty=0  \,, \label{WINF}
\end{equation}
is satisfied, in agreement with Eq. \eqref{f(R,T):int_mathcalP}. Although we are using global K-monopoles as a particle toy model, the result is much more general. For example, in GR, atomic nuclei are also required to satisfy the standard von Laue condition, albeit exhibiting a different pressure profile \cite{Polyakov:2018zvc, Shanahan:2018nnv, Burkert:2018bqq} (as is also the case for nuclei with $B>1$ in the Skyrme model \cite{GarciaMartin-Caro:2023toa}). Despite these differences in detail, realistic particles that admit a soliton-like description should obey the modified von Laue condition in $R+F(T)$ gravity, given in Eq.~\eqref{von laue}, independently of the details of their composition.

\subsection{Matter on-shell Lagrangian}

To assess the on-shell relation between the matter Lagrangian and the trace of the energy--momentum tensor, we compare their spatial averages, considering in particular the $r \to \infty$ limit.

Taking into account that $\mathcal{L}_{\rm m}=-\rho$ and $T=-\rho+3p$, one finds
\begin{equation}
\frac{\langle \mathcal{L}_{\rm m} \rangle_r}{\langle T \rangle_r}=\frac{\langle\rho\rangle_r}{-\langle\rho\rangle_r+3\langle p\rangle_r}=\frac{1}{1-3 w_r} \,.
\end{equation}
Since, in general, $w_\infty \equiv \langle p\rangle_\infty/\langle\rho\rangle_\infty \neq 0$, the global spatial averages of $\mathcal{L}_{\rm m}$ and $T$ do not, in general, coincide.

On the other hand, 
\begin{equation}
\frac{\langle \mathscr{L}_{\rm m} \rangle_r}{\langle \mathscr{T} \rangle_r}=\frac{-\langle\varrho\rangle_r}{-\langle\varrho\rangle_r+3\langle\mathscr P\rangle_r}=\frac{1}{1-3{\mathcal W}_r} \,.
\end{equation}
Given that ${\mathcal W}_\infty =0$, it follows that
\begin{equation}
\frac{\langle \mathscr{L}_{\rm m} \rangle_\infty}{\langle \mathscr{T} \rangle_\infty}=1\,.
\end{equation}

Therefore, the equality $\mathscr{L}_{\rm m}=\mathscr{T}$ holds on average within a global K-monopole.

\subsection{$F(\langle T\rangle_r)$ and $\langle F(T)\rangle_r$ for global K-monopoles}

\begin{figure}[t!]
\centering
\includegraphics[width=\columnwidth]{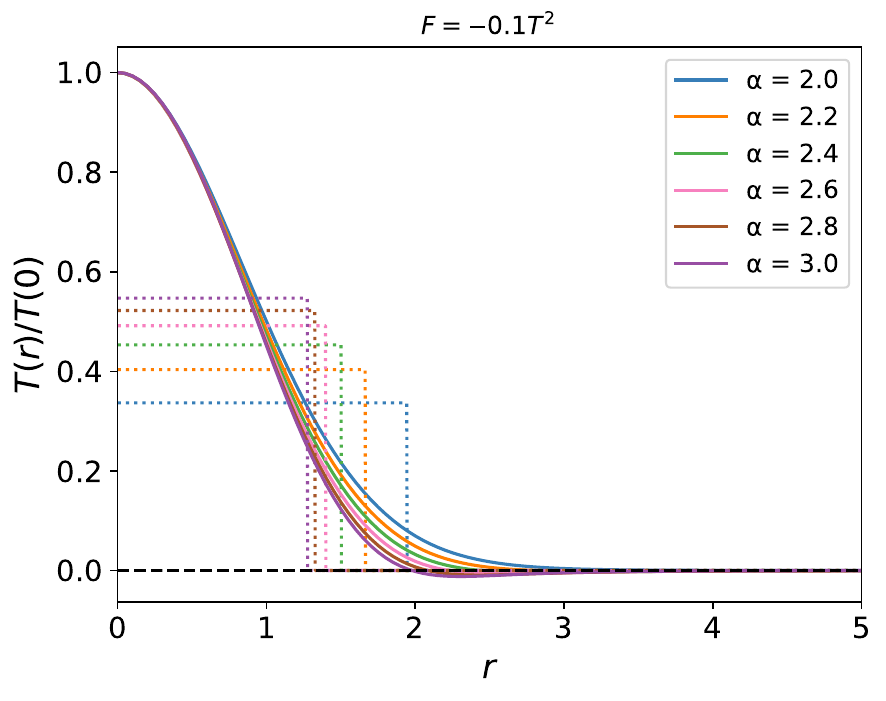}
\caption{The solid lines show $T/T(0)$ as a function of the distance from the monopole center $r$, for various values of $\alpha$ and $F(T) = -0.1\,T^2$ (the dashed line corresponds to 
$T/T(0) = 0$). The dotted lines, representing the ratio $T_{\rm TH}/T(0)$ obtained considering the simplified top-hat profile for the radial dependence of $T$, provide only a rough representation of the actual behavior of $T$. 
\label{fig4}
}
\end{figure}

In this subsection, we compare the spatial average of $F$ with $F$ evaluated at the spatial average of $T$. To this end, we compute the ratio
$\langle F(T)\rangle_r / F(\langle T\rangle_r)$ using global K-monopoles as a
particle toy model. We also consider an approximation based on the following simplified top-hat profile for the radial dependence of the trace $T$ of the energy-momentum tensor,
\begin{equation}
T_{\rm TH}(r)=  \begin{cases}
        T_*\ \ \text{, } r<r_* \,,\\
        0\, \ \ \text{, } r>r_*\,,
       \end{cases}
\end{equation}
where the constant $T_*$ is equal to
\begin{equation}
T_* = \frac{1}{r_*^3}\lim_{r \to \infty} (r^3 \langle T \rangle_r)\,,
\label{T*c}
\end{equation}
and $r_*$ is defined by the condition
\begin{eqnarray}
&&\lim_{r \to \infty} \frac{\langle F(T)\rangle_r F(\langle T_{\rm TH}\rangle_r)}{F(\langle T\rangle_r) \langle F(T_{\rm TH})\rangle_r} \nonumber \\
&=& \lim_{r \to \infty} \frac{\langle F(T)\rangle_r F(T_*{r_*^3}/{r^3})}{F(\langle T\rangle_r) F(T_*){r_*^3}/{r^3}}= 1\,.
\label{r*c}
\end{eqnarray}
Figure \ref{fig4} shows that the profile of $T_*/T(0)$ (represented by the dotted lines),  obtained for various values of $\alpha$ assuming that $F(T)=-0.1T^2$, provides only a rough representation of the actual behavior of $T/T(0)$ (represented by the solid lines).

Equations~\eqref{T*c} and~\eqref{r*c} ensure that not only the volume integral of the trace of the energy-momentum tensor matches that of the original model as $r \to \infty$, but also that the ratio between $\langle F(T)\rangle_r / F(\langle T\rangle_r)$ and $\langle F(T_{\rm TH})\rangle_r / F(\langle T_{\rm TH}\rangle_r)$ asymptotically converges to unity in this limit. It follows that the parameters $r_*$ and $T_*$ are inherently model dependent. Nevertheless, if $T$ decreases sufficiently rapidly with $r$, $r_*$ generally provides an excellent estimate of the physical size of the particle.

\begin{figure}[t!]
\centering
\includegraphics[width=\columnwidth]{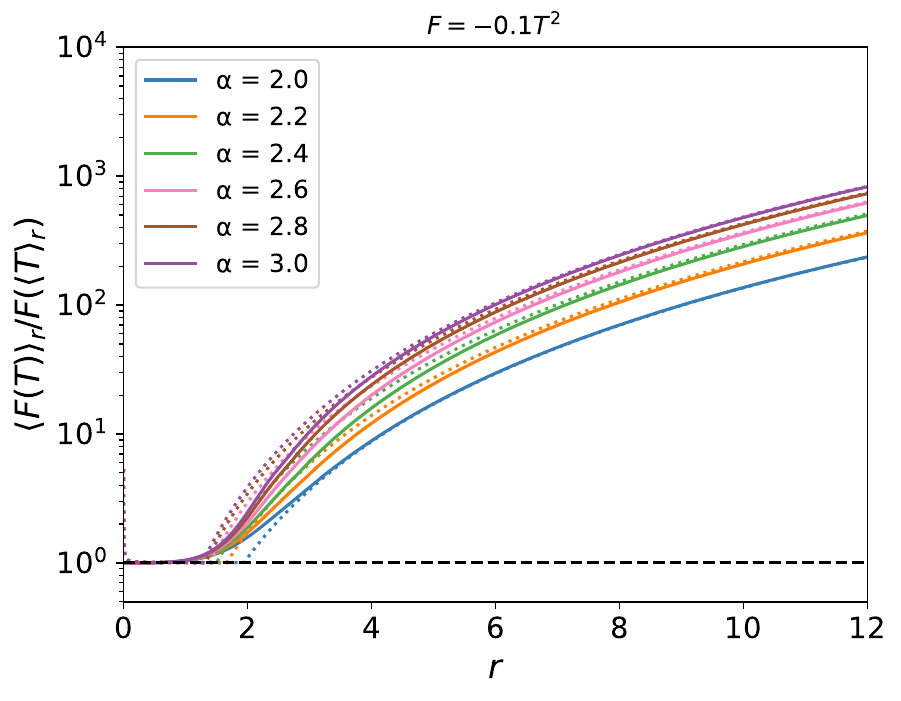}
\caption{The solid lines display the behavior of 
$\langle F(T)\rangle_r / F(\langle T\rangle_r)$ as a function of the distance from the 
monopole center $r$, for several values of $\alpha$ and $F(T) = -0.1\,T^2$ 
(the dashed black line corresponds to 
$\langle F(T)\rangle_r / F(\langle T\rangle_r) = 1$). 
The dotted lines represent the behavior of 
$\langle F(T_{\rm TH})\rangle_r / F(\langle T_{\rm TH}\rangle_r)$ obtained using the analytical 
approximation, which shows excellent agreement with the numerical results at large $r$.
\label{fig5}
}
\end{figure}

Although Eq. \eqref{r*c} is strictly valid in the $r \to \infty$ limit, it is expected to effectively hold for $r \gg r_*$. Therefore, in this regime
\begin{equation}
\frac{\langle F(T)\rangle_r}{F(\langle T\rangle_r)} \sim \frac{\langle F(T_{\rm TH})\rangle_r}{F(\langle T_{\rm TH}\rangle_r)}  = \frac{F(T_*)}{F(T_*{r_*^3}/{r^3}) } \left(\frac{r_*}{r}\right)^3 \label{ratio}\,.
\end{equation}
Notice that, except if $F(T) \propto T$, this ratio depends on $r$ and can be significantly different from unity. For example, if $F(T)=\epsilon |T|^{\beta}$ then
\begin{equation}
\frac{\langle F(T)\rangle_r}{F(\langle T\rangle_r)} \sim \left(\frac{r}{r_*}\right)^{3 (\beta-1)} \label{ratio1}\,,
\end{equation}
for $r \gg r_*$.

The solid lines in Fig.~\ref{fig5} display the ratio 
$\langle F(T)\rangle_r / F(\langle T\rangle_r)$ as a function of the distance from the monopole center $r$, for several values of $\alpha$ and for the model $F(T) = -0.1\, T^2$. Figure~\ref{fig5} shows that, at large $r$, this ratio closely follows the approximation obtained using a simplified top-hat profile for the radial dependence of $T$, as given in Eqs.~\eqref{ratio} and~\eqref{ratio1}. Notably, although the latter depends on $r_*$ it is otherwise independent of the particle composition.

\section{Cosmological implications \label{Sec_fluids}}

On cosmological scales, the energy-momentum content of the Universe is typically modeled as a collection of perfect fluids. This description provides an effective characterization of the large-scale dynamics of the Universe by capturing the collective behavior of large particle ensembles. Assuming large-scale homogeneity and isotropy, and restricting to the spatially flat case, the spacetime geometry is described by the flat FLRW metric, with line element
\begin{equation}
ds^2 = -dt^2 + d\vec{r} \cdot d\vec{r}
      = -dt^2 + a^2(t)\, d\vec{x} \cdot d\vec{x} \, ,
\end{equation}
where $t$ denotes physical time, $a(t)$ is the scale factor, $\vec{x}$ are comoving Cartesian coordinates, and $d\vec{r} \equiv a(t)\, d\vec{x}$.

For an ideal gas, any coarse-grained (fluid-level) variable $ \overline{Q} (t,\vec{x})$ is obtained from the underlying particle contributions $Q_i$ as
\begin{equation}
\overline{Q}(t,\vec{x})
= \frac{1}{V_{\rm c}}
\int_{V_{\rm c}} \sum_{i=1}^N Q_i(t,\vec q\,) \, d^3q \, ,
\label{A_fluid}
\end{equation}
where $V_{\rm c}$ is a small comoving volume centered at $\vec{x}$ and containing a large number of particles $N$. Here, we implicitly adopt the standard approximation in which geometrical backreaction is neglected, assuming an exactly FLRW spacetime when performing cosmological averaging over spatial hypersurfaces.

Consider a dust fluid of stable cold dark matter particles with fixed proper mass $m$ and proper modified mass $\mathcal{M}$, comoving with the Hubble flow. In an FLRW universe, the characteristic interparticle distance,
\begin{equation}
\ell \equiv (\overline{n})^{-1/3}\,,
\end{equation}
scales as 
\begin{equation}
\ell \propto a\,,
\end{equation}
thus implying that
\begin{equation}
\rho = m\,\overline{n} \propto a^{-3}\,, \qquad \varrho = \mathscr{M}\,\overline{n} \propto a^{-3}\,.
\end{equation}
This behaviour is fully consistent with the evolution of $\overline{\varrho} = \mathscr{M}\,\overline{n}$ obtained from Eq.~\eqref{Tmodcons}, using the modified energy--momentum tensor defined in Eq.~\eqref{mathcalT2}, which yields
\begin{equation}
\dot{\overline{\varrho}} + 3H\left(\overline{\varrho}+\overline{\mathscr{P}}\right)=0\,.
\label{continuity2}
\end{equation}
Imposing $\overline{\mathscr{P}} = 0$, in agreement with the modified von~Laue condition, immediately gives
\begin{equation}
\overline{\varrho}=\mathscr{M} \, \overline{n} \ \propto \ a^{-3}\,. \label{varrhoa}
\end{equation}
This result hinges on the fact that in $R+F(T)$ gravity it is the proper modified pressure
\begin{equation}
\overline{\mathscr{P}} = \overline{p + F + 2F_{,T}(p +{\mathbb{T}}^i_{\ i}/3)}
\end{equation}
of the fluid, which is required to vanish, rather than its proper pressure $\overline{p}$ (as we have shown in the previous section, the latter is generally nonzero). This subtlety is often overlooked in the literature, where $\overline{p}$ is typically assumed to vanish for dust in $f(R,T)$ gravity (see, e.g., \cite{Alvarenga:2012bt,Alvarenga:2013syu,Zaregonbadi:2016xna,Velten:2017hhf,Rudra:2020nxk, Haghani:2023uad,Solanke:2023aty}). 

Assuming that $\ell$ is much larger than the radius of the dust particles, the macroscopic quantity $\overline{Q}$ can be approximated as 
\begin{equation}
\overline{Q} = \langle Q \rangle_\ell \,.
\end{equation}
Here, the right-hand side is obtained by considering a single particle and performing a volume average within a sphere of radius $r=\ell$, as in Eq.~\eqref{average_r} of the previous section.

A common but generally unjustified simplification is to assume that $\overline{F}=F(\overline{T})$. According to Eq.~\eqref{A_fluid}, large-scale (coarse-grained) variables must be defined as spatial averages of the corresponding microscopic counterparts, implying that the correct quantity at the fluid-level is $\overline{F}$ rather than $F(\overline{T})$. This distinction becomes particularly important when $F(T)$ depends nonlinearly on $T$ (e.g., if $F(T)=\epsilon |T|^\beta$ with $\beta \neq 1$), in which case $\overline{F} \neq F(\overline{T})$. For $F(T) = \epsilon |T|^\beta$, the ratio between the averaged quantities $\overline{F}$ and $F(\overline{T})$ is
\begin{align}
\mathfrak{f} \equiv \frac{\overline{F}}{F(\overline{T})}
&\sim \frac{\frac{4\pi \epsilon}{\ell^3} \int_0^\ell |T|^\beta r^2 \, dr}
       {\epsilon \left( \frac{4\pi}{\ell^3} \int_0^\ell |T| r^2 \, dr \right)^\beta} \nonumber\\
&= \left( \frac{4\pi}{\ell^3} \right)^{1-\beta}
   \frac{\int_0^{r_*} |T_*|^\beta r^2 \, dr}
        {\left( \int_0^{r_*} |T_*| r^2 \, dr \right)^\beta} \nonumber\\
&= \left( \frac{4\pi r_*^3}{3 \ell^3} \right)^{1-\beta}
  = \left( \frac{{r}_+}{\ell} \right)^{3(1-\beta)} \, ,
\label{ratio_lambda}
\end{align}
where $r_*$ is the particle radius defined in the previous section and $r_+=(4\pi/3)^{1/3}r_*$. Thus, $\mathfrak{f}$ generally deviates from unity as $\ell$ increases relative to the particle radius (except when $\beta = 1$). In what follows, we show that assuming a unit value for $\mathfrak{f}$ can significantly affect the inferred cosmological evolution.

For a dust fluid in an FLRW background, the $00$ component of Eq.~\eqref{Eqm_g_2} becomes
\begin{equation}
3 H^2 = \frac{1}{2} \overline{\varrho} = \frac{1}{2} \left(\overline{\rho}-  \overline{F} \right)
= \frac{1}{2} \Bigl(\overline{\rho} - \mathfrak{f}\, F(\overline{T})\Bigr)\,,
\label{Friedmann_Final_beta}
\end{equation}
where $\overline{T} = -\overline{\rho} + 3 \overline{p}$, $H = \dot a /a$, and a dot denotes a derivative with respect to the physical time $t$. Assuming that $F(T)=\epsilon |T|^\beta$, an exact expression for $\mathfrak{f}$ can be obtained by noting that
\begin{eqnarray}
\overline{F} &=& \overline{\rho}-\overline{\varrho}= \overline{n}\left(m-\mathscr{M}\right)= \overline{\rho} \left(1-\frac{\mathscr{M}}{m}\right),\\
F(\overline{T}) &=& \epsilon\,|-1+3 w_\infty|^\beta\,\overline{\rho}^{\,\beta}.
\end{eqnarray}
As a result,
\begin{equation}
\mathfrak{f}
\equiv
\frac{\overline{F}}{F(\overline{T})}
=
\epsilon^{-1}
\left(1-\frac{\mathscr{M}}{m}\right)
|-1+3 w_\infty|^{-\beta}
\overline{\rho}^{\,1-\beta}\,.
\end{equation}
Hence, for $\beta \neq 1$,
\begin{equation}
\mathfrak{f} \propto a^{3(\beta-1)}
\end{equation}
exhibits an explicit dependence on the scale factor $a$ and can deviate significantly from unity, a feature also neglected in previous studies \cite{Alvarenga:2013syu,Zaregonbadi:2016xna,Velten:2017hhf,Rudra:2020nxk,Haghani:2023uad}, which implicitly assumed a unit $\mathfrak{f}$. Therefore, incorrectly setting $\mathfrak{f}$ to unity alters Eq.~\eqref{Friedmann_Final_beta}, resulting in inconsistent cosmological dynamics and a distorted cosmic evolution.

Equations~\eqref{varrhoa} and~\eqref{Friedmann_Final_beta} imply that
\begin{equation}
H^2 \propto a^{-3}\,,
\end{equation}
which leads to
\begin{equation}
a \propto t^{2/3}\,,
\end{equation}
characteristic of the expansion in GR of a flat, matter-dominated FLRW universe. This analysis emphasizes that consistent volume averaging, together with the modified von~Laue condition, is essential to recover the correct cosmological evolution. Incorrectly identifying $\overline{F}$ with $F(\overline{T})$ or assuming $\overline{p}=0$ can produce spurious deviations from the standard dust-like behavior, potentially leading to misleading conclusions regarding the expansion history and the evolution of the energy density in this class of models.

The preservation of the standard matter-dominated evolution of a universe filled with dust in $R + F(T)$ gravity, independently of the specific form of $F(T)$, follows directly from two key features. The first is a distinctive property of $R+F(T)$ gravity, namely its equivalence to GR coupled to a modified matter Lagrangian. The second stems from the fact that the corresponding modified energy-momentum tensor satisfies the modified von Laue condition. Thus, the preservation of the standard cosmological evolution naturally extends to the case of moving particles, indicating that, in $R+F(T)$ gravity, the evolution of the Universe during both the matter- and radiation-dominated eras remains effectively unchanged compared to GR.

Although much of the above discussion has focused on pressureless dust in $R+F(T)$ gravity, the generalization to $f_1(R)+f_2(T)$ gravity is straightforward. More generally, several of the issues identified here are expected to extend to other matter sources and to a much broader class of nonminimally coupled gravity theories. In particular, in $f(R,\mathcal{L}_{\rm m},T)$ theories with nonlinear dependence on $\mathcal{L}_{\rm m}$ or $T$, or more generally in any framework where the cosmological field equations contain nonlinear functions of the averaged fluid energy density, an inconsistent averaging procedure can similarly lead to misleading cosmological predictions.

\section{Conclusions \label{concl}}

We investigated the cosmological averaging problem in theories of gravity with nonminimal couplings between matter and geometry, focusing on its impact on the large-scale evolution inferred from homogeneous cosmological models. We have shown that while averaging issues are present even in GR, they can become particularly significant in theories with nonlinear matter-gravity interactions.

Using $R+F(T)$ gravity as a well controlled minimal framework and global K-monopoles as a particle proxy, we explored the challenges associated with cosmological averaging for a fluid composed of dust, clarifying several common misconceptions in the literature. In particular, we have shown that homogeneous cosmological models constructed under the assumption that $\overline{F}=F(\overline{T})$ generally fail to reproduce the correct large-scale dynamics of an inhomogeneous universe. We have further demonstrated that, contrary to GR, dust is generally not pressureless in these theories. 

Although our discussion has focused on the specific case of $R+F(T)$ gravity, the underlying analysis is not restricted to this particular model. Rather, it applies more generally to theories of gravity featuring nonminimal couplings between matter and geometry, provided the standard assumption, widely adopted in a cosmological context, that spacetime can be well approximated by an FLRW metric on all relevant scales is maintained. Whether this assumption remains valid in such theories is ultimately a question for a more complete theoretical framework, particularly in light of the instabilities that have been reported in significant parts of the parameter space of nonminimally coupled gravity. Should these issues be resolved, the averaging problem investigated in this work would persist and must be addressed consistently.

In addition, we have deliberately not incorporated the stringent experimental constraints at redshift zero that severely restrict the functional form of $F(T)$. Our aim has not been to assess the phenomenological viability of these theories, but rather to use $R+F(T)$ gravity as a controlled setting to investigate the robustness of homogenization procedures themselves, independently of the current experimental and observational limits.

Finally, an important ingredient that deserves further investigation is the role of geometrical backreaction. Although its effects within GR are commonly assumed to be small, this assumption does not necessarily extend to modified gravity theories. Indeed, even in the context of $f(R)$ gravity, geometrical backreaction has already been shown to be relevant, in particular if  a large fraction of dark matter turns out to be black holes \cite{Cano:2024anc}. A consistent treatment of geometrical backreaction is therefore essential and should be incorporated in future studies of averaging in nonminimally coupled gravity theories.

\acknowledgments

We thank our colleagues of the Cosmology group at Instituto de Astrofísica e Ciências do Espaço for enlightening discussions and acknowledge the support by Fundação para a Ciência e a Tecnologia (FCT) under the research grant UID/04434/2025 (DOI 10.54499/UID/04434/2025). S. R. P. also acknowledges the support by Fundação para a Ciência e a Tecnologia (FCT) through the grant No. 2025.03891.BD.
\bibliography{article}

\end{document}